\documentclass[9pt,twocolumn]{article}
\ifdefined\XeTeXversion\else\pdfoutput=1\fi

\usepackage[margin=0.75in,columnsep=0.25in]{geometry}
\usepackage[T1]{fontenc}
\usepackage{lmodern}

\usepackage{balance}

\usepackage{amsmath}
\usepackage{amssymb}

\usepackage{booktabs}
\usepackage{array}

\usepackage{enumitem}

\usepackage{tikz}
\usetikzlibrary{arrows.meta, positioning, fit, calc, backgrounds, shapes, decorations.pathreplacing}
\usepackage{pgfplots}
\pgfplotsset{compat=1.18}

\usepackage[linesnumbered,ruled,vlined]{algorithm2e}

\usepackage{microtype}

\usepackage{xcolor}

\usepackage{graphicx}

\usepackage{caption}
\usepackage{subcaption}
\usepackage{float}

\usepackage{url}
\expandafter\def\expandafter\UrlBreaks\expandafter{\UrlBreaks%
  \do\-\do\0\do\1\do\2\do\3\do\4\do\5\do\6\do\7\do\8\do\9}

\usepackage[numbers]{natbib}

\usepackage[hidelinks]{hyperref}

\usepackage{titlesec}
\titleformat{\section}{\normalsize\bfseries}{\thesection}{0.5em}{}
\titleformat{\subsection}{\small\bfseries}{\thesubsection}{0.5em}{}
\titlespacing*{\section}{0pt}{\baselineskip}{0.5\baselineskip}
\titlespacing*{\subsection}{0pt}{0.5\baselineskip}{0.25\baselineskip}

\title{\textbf{Lifecycle-Aware Archival for Asymmetric Financial\\
Datasets: A Production Study}}

\author{
  Tulika Manek \\
  \textit{Razorpay}
}

\date{}

\begin{document}

\twocolumn[\maketitle]

\noindent\textbf{Abstract.}
Large-scale financial transaction databases face a fundamental tension between
operational freshness requirements and storage efficiency. We present the
design, implementation, and production evaluation of a lifecycle-aware archival
system for a financial transaction service at Razorpay managing billions of records
on PostgreSQL Aurora (version~14), occupying tens of terabytes of storage and sustaining peak
write throughput in the thousands of TPS.
We make two contributions. First, we analytically characterize the
\emph{Celebrity Partition Problem}: lifecycle-state partitioning concentrates all
operationally active rows in a single default partition, incurring
$O(N{\times}M)$ planning overhead, $O(N)$ execution I/O regression, and write
amplification on every state transition. Second, we
present an ID-monotonicity-based deduplication technique that exploits temporal
encoding in monotonically increasing ID schemes (Snowflake IDs, ULIDs, and
equivalent) to route only potentially-archived re-inserts to a warm database
lookup, requiring no Bloom filters, idempotency tables, or external dependencies.
We report production results from a fully deployed system: 95\% hot storage size reduction, 53\% reduction in monthly database infrastructure cost,
${\sim}60\%$ reduction in service-level p99 processing latency, 51~percentage-point
reduction in writer CPU utilization, and sustained operation at peak write
TPS - without schema changes to the primary transaction table.

\section{Introduction}
\label{sec:intro}

Financial transaction systems impose a constraint absent from general-purpose
databases: a record cannot be archived until its associated financial event -
settlement, in our system - has been fully confirmed, regardless of how old it
is. There is therefore no hard upper bound
on the dataset's size.

At Razorpay, we operate one such system, which has accumulated over a decade of
transaction records. The transactions table supports
the full spectrum of workloads - high-TPS writes, batch
scans, reconciliation, and audit - and carries seven secondary indexes to serve
these access patterns. Table~1 (Section~3) characterizes the system's scale in
full.

The scale of this dataset creates two compounding operational problems. First,
PostgreSQL implements row updates as a delete-plus-insert under its MVCC model,
leaving dead tuples that must be reclaimed by the autovacuum daemon. With over
85\% of new transactions reaching their terminal status within their creation day,
dead tuples accumulate continuously on a multi-terabyte heap - inflating
storage and degrading scans, which must traverse dead
rows to find live ones. Second, periodic batch processing must scan all pending
transactions for a given merchant to compute the amount due. Against a
table of this scale with seven secondary indexes, such scans can run for several
minutes for high-volume merchants, with service-level p99 latency reaching
approximately 250~seconds.
Temporary mitigations - query tuning,
infrastructure scaling, application-side timeout handling - provide short-term
relief but are fundamentally insufficient: the dataset grows by hundreds of millions of rows per month, so any fixed-capacity intervention is eventually
overtaken by volume growth.

\subsection{The Lifecycle Asymmetry}

The key structural insight that drives our design is a sharp lifecycle
asymmetry intrinsic to such transaction data. Once a transaction is settled it
reaches its terminal status: it is never written to again and becomes
immutable, read-only historical data. Active (unsettled) transactions, by
contrast, are write-intensive: they undergo state transitions and updates
throughout their lifecycle - the source of the dead-tuple churn described
above.

At Razorpay, approximately 94\% of transactions reach
terminal status within two months and become
permanently immutable. Only ${\sim}6\%$ remain
active - the subset that receives all writes and scans. This asymmetry suggests
a clear archival target: if terminal transactions could be removed from hot
storage, the active dataset would shrink by ${\sim}94\%$, dramatically reducing
both storage cost and scan scope, without losing point-lookup latency,
range-scan capability, or operational SLAs. However, realizing this in practice requires
solving two non-obvious problems that form the technical contributions of this
paper.

\subsection{Why This Is Hard}

\textbf{The partitioning trap.} An obvious response to a large
table with a clear archival axis is to partition on that axis. For settlement
data, \texttt{settled\_at} - the timestamp of the terminal settlement
event - appears to be the obvious partition key: old
partitions could then be dropped as an $O(1)$ metadata operation. We show in
Section~4 that this intuition fails for lifecycle-asymmetric datasets in a way
that is non-obvious until it is too late to reverse - a failure mode we refer to
as the Celebrity Partition Problem (Section~4).

\textbf{The deduplication problem.} Removing records from the primary database
breaks the uniqueness guarantee of the primary key. If a record is archived and
later re-inserted by a replay or recovery process, the insert will succeed,
creating a duplicate. In a financial system, duplicate records can trigger
duplicate settlements, causing monetary loss, discrepancies, and reconciliation issues. Maintaining a separate
deduplication index over billions of IDs is itself a storage-scaling problem.
We show in Section~6 that a structural property of monotonically generated
IDs - that their prefix encodes creation time - allows duplicate detection
without any additional storage, using only a conditional warm-database lookup
for IDs older than the archival window.

\section{Background}
\label{sec:background}

\subsection{PostgreSQL MVCC and Autovacuum}

PostgreSQL's MVCC model creates a new physical tuple on every \texttt{UPDATE}
and marks the old version as dead~\cite{PostgreSQL14}. Dead tuples are reclaimed
by autovacuum, whose throughput is bounded by configurable I/O cost limits. When
dead tuple creation persistently outpaces reclamation, the heap grows (table bloat)
and scan performance degrades as queries traverse dead rows.

\subsection{PostgreSQL Partitioning}

Declarative partitioning~\cite{PostgreSQL14} prunes irrelevant partitions when
queries constrain the partition key. Rows with a NULL key are routed to a default
partition that can never be pruned. Each partition carries its own copy of every
index ($N{\times}M$ total). For queries that do not constrain the partition
key, this imposes $O(N{\times}M)$ planning overhead and up to $O(N)$ execution
I/O. Cross-partition row movement (on key-value change) incurs
$2{\times}M$ index operations per update.

\subsection{Transaction Lifecycle: Active and Terminal Phases}

A transaction record has two phases with fundamentally different
storage requirements. In the active phase, the record
receives state-changing updates throughout its lifetime. In the terminal
phase, the terminal timestamp (\texttt{settled\_at} in our system) is recorded
once and the record becomes permanently immutable.

Critically, terminal status is independent of record age. A transaction may
remain active indefinitely due to account configuration, risk holds, or
policy rules, while a transaction created the previous day may already have
reached terminal status. This creates an archival challenge distinct from time-based archival:
the relevant axis is lifecycle status, not creation time - the natural partition
key is a column that is NULL for all operationally active rows.

\section{System Characteristics and Data Profile}
\label{sec:system}

\subsection{Infrastructure}

The service runs on PostgreSQL Aurora (version~14) on AWS.
PostgreSQL Aurora uses a shared distributed storage layer across the
writer and all reader instances in a cluster: all instances read from and write
to the same storage volume, and storage auto-scales in 10~GB increments. This
architecture means that storage consumed by table bloat or large indexes is
shared across the entire cluster - there is no per-instance storage to optimize
independently.

All database clusters (hot and warm) reside within the regulated geography,
consistent with applicable data-residency requirements. The system operates
within a PCI~DSS-scoped environment; archival does not alter the
cardholder-data boundary, as both hot and warm clusters remain within the
existing compliance perimeter.

A separate warm database cluster - an I/O-optimized Aurora instance - is
maintained for historical data queries that the primary cluster cannot serve
without impacting write throughput. The warm cluster is populated via
change-data-capture (CDC) replication from the primary, receiving all writes
with a short replication lag.

\subsection{Data Characteristics}

Table~\ref{tab:characteristics} summarizes the pre-archival state of the transactions table. A notable
characteristic is that index storage exceeds table data storage - an inversion
driven by seven secondary indexes on a heavily updated table.
Each update generates dead index entries alongside dead heap tuples. With seven indexes,
total index dead-entry accumulation is roughly seven times that of the heap,
and index pages require a separate vacuum pass not covered by the heap scan,
compounding the reclamation burden on autovacuum.

With over 85\% of incoming transactions settling within the same day of arrival,
settlement updates generate dead tuples at nearly the full insert rate. The peak
dead tuple creation rate reaches thousands per second (peak write throughput
$\times$ 85\% same-day settlement rate). At the
observed average monthly volume of hundreds of millions of new transactions, the
average dead tuple creation rate is hundreds of millions per month.
The distinction matters: peak dead tuple creation rate governs the
immediate I/O pressure on autovacuum during high-traffic periods, while the
monthly average governs the long-run storage growth rate. Without archival, the
table grows continuously - live rows from new transactions plus accumulated dead
tuples from updates. Absent intervention, storage and cost would
roughly double within a few years at observed rates, making unbounded table growth
both a performance risk and a direct cost escalation risk.

\begin{table}[!t]
\centering\small
\caption{Transactions table characteristics prior to archival.}
\label{tab:characteristics}
\begin{tabular}{@{}ll@{}}
\toprule
\textbf{Metric} & \textbf{Value} \\
\midrule
Total rows                  & billions \\
Table data size             & tens of TB \\
Index size                  & exceeds table data size \\
Total (table + index)       & tens of TB \\
Total DB size               & tens of TB \\
Secondary indexes           & 7 \\
Peak write TPS              & thousands \\
Settled transactions        & $\sim$94\% of total rows \\
Unsettled transactions      & $\sim$6\% of total rows \\
Monthly new transactions    & hundreds of millions \\
\bottomrule
\end{tabular}
\end{table}

\section{Design Space Evaluation}
\label{sec:design}

We evaluated three candidate approaches to reducing the active dataset.
Approaches A and B were evaluated analytically against documented PostgreSQL
behavior and operational constraints; and Approach~C was deployed. The
comparisons below are therefore design-time predictions for A and B, and
measured outcomes for C (Section~7).
Table~\ref{tab:comparison} summarizes the trade-offs; the subsections below
explain the reasoning behind each row. The approach selected for implementation
(Approach~C) is described in detail in Section~5.

\subsection{Approach A: Two-Table Split (Settled and Unsettled)}

The most direct interpretation of the lifecycle asymmetry is to maintain two
separate tables: an operational table containing only active (unsettled)
transactions and an archival table containing settled transactions. When a transaction is
settled, it is deleted from the operational table and inserted into the archival
table. The archival table is then periodically purged or dropped.

\textbf{Write amplification.} Every settlement event requires: (1) an INSERT
into the archival table carrying the settlement metadata, and (2) a DELETE from
the operational table. Each operation updates all $M$ secondary indexes on its
respective table, effectively doubling the index write cost of each settlement
compared to a single-table update. The cross-table movement also requires
distributed transaction semantics or careful idempotency handling to prevent
data loss on partial failure.

\textbf{Operational complexity.} Queries spanning both settled and unsettled
data - merchant dashboards, reconciliation jobs, audit queries - must fan out to
both tables and merge results in the application layer. Any query that does not
know in advance whether a record is settled or unsettled must query both tables.
This doubles read load for ambiguous queries and complicates the application
layer with table-aware routing logic.

\textbf{Migration cost.} The initial migration requires moving the full settled
backlog (${\sim}94\%$ of all rows) into the archival table with all associated
index updates. At production write rates, this operation would take days and
require the system to operate in a degraded or frozen-writes state during the
transition.

\subsection{Approach B: Partition on \texttt{settled\_at}  -  The Celebrity Partition Problem}

The second approach is to partition the transactions table on the
\texttt{settled\_at} column using PostgreSQL's range partitioning. Monthly
partitions contain settled transactions for each time period; old partitions can
be detached and dropped as an $O(1)$ metadata operation. This approach is
appealing because partition drops impose no I/O, no write amplification, and no
application-level complexity beyond routing.

We note that partitioning on a sparse or nullable column is a known anti-pattern;
what is less documented is the specific interaction with PostgreSQL's default
partition routing, the resulting $O(N{\times}M)$ planning overhead and $O(N)$
execution I/O regression, and the write amplification on every lifecycle state
transition - the combination of which makes this approach strictly worse than no
partitioning for queries that do not filter on the partition key.

\textbf{The Celebrity Partition Problem.} In PostgreSQL range partitioning, rows
whose partition key is NULL are routed to the default partition - a catch-all
partition that is never pruned by queries that do not filter on the partition
key. Because \texttt{settled\_at} is NULL for all unsettled transactions by
definition (a transaction has no settlement timestamp until it is settled), every
unsettled row lands in the default partition. The default partition therefore
contains the entire operationally active dataset - ${\sim}6\%$ of total rows by
count but 100\% of writes and settlement-scan reads.

This is the Celebrity Partition Problem. The name reflects the access
pattern: one partition among many attracts nearly all of the system's
attention, as a celebrity does in a crowd - the lifecycle-state analogue of
the hot-key problem in sharded systems. When a partition scheme uses a
lifecycle-state column as the partition key, the operationally active subset
has a NULL key value, causing all active rows to concentrate in the default
partition. The default partition becomes disproportionately large and
disproportionately accessed - behaviorally equivalent to the original
unpartitioned table. The partitioned settled data provides no benefit to queries that do not filter
on the partition key, while the overhead of partitioning actively degrades
such queries.

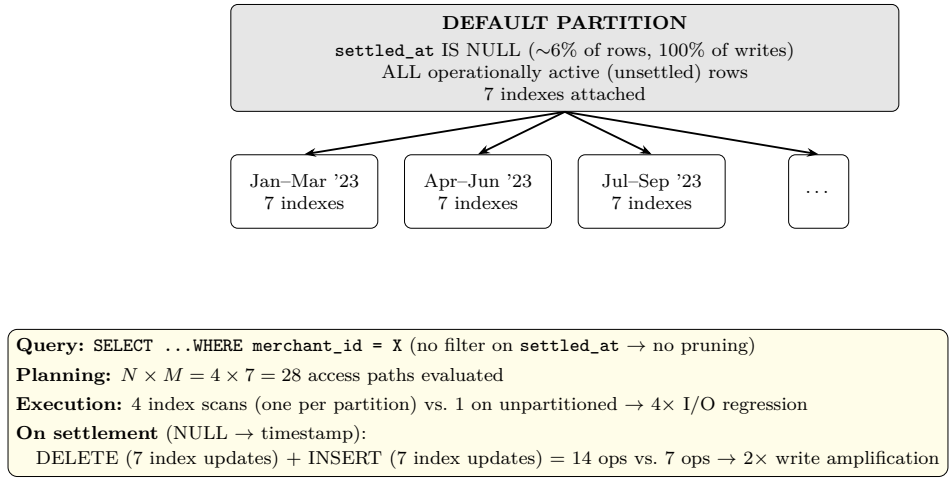
\begin{figure*}[!t]
\centering
\resizebox{0.7\textwidth}{!}{%
\begin{tikzpicture}[
  box/.style={draw, rounded corners=3pt, align=center, font=\footnotesize},
  arr/.style={-{Stealth[length=5pt]}, thick}
]

\node[box, fill=gray!20, minimum width=10cm, minimum height=1.6cm] (def) at (0,0)
  {\textbf{DEFAULT PARTITION}\\[2pt]
   \texttt{settled\_at} IS NULL ($\sim$6\% of rows, 100\% of writes)\\
   ALL operationally active (unsettled) rows\\
   7 indexes attached};

\node[box, minimum width=2.2cm, minimum height=1.1cm]
  (p1) at (-3.9,-2.0) {Jan--Mar '23\\7 indexes};
\node[box, minimum width=2.2cm, minimum height=1.1cm]
  (p2) at (-1.3,-2.0) {Apr--Jun '23\\7 indexes};
\node[box, minimum width=2.2cm, minimum height=1.1cm]
  (p3) at (1.3,-2.0)  {Jul--Sep '23\\7 indexes};
\node[box, minimum width=0.9cm, minimum height=1.1cm]
  (pe) at (3.8,-2.0)  {\ldots};

\foreach \child in {p1,p2,p3,pe}
  \draw[arr] (def.south) -- (\child.north);

\node[draw, rounded corners, fill=yellow!10,
      below=1.5cm of p2, anchor=north,
      minimum width=10.0cm, align=left, font=\footnotesize] (ann) {
  \textbf{Query:} \texttt{SELECT \ldots WHERE merchant\_id = X}
  (no filter on \texttt{settled\_at} $\to$ no pruning)\\[3pt]
  \textbf{Planning:} $N \times M = 4 \times 7 = 28$ access paths evaluated\\[3pt]
  \textbf{Execution:} 4 index scans (one per partition) vs.\ 1 on unpartitioned
  $\to$ $4\times$ I/O regression\\[3pt]
  \textbf{On settlement} (NULL $\to$ timestamp):\\[1pt]
  \quad DELETE (7 index updates) + INSERT (7 index updates) = 14 ops vs.\ 7 ops $\to$ $2\times$ write amplification
};

\end{tikzpicture}%
}
\caption{The Celebrity Partition Problem: all operationally active rows
concentrate in the default partition because \texttt{settled\_at} is NULL until
settlement. Queries on \texttt{merchant\_id} cannot prune any partition, incurring
$O(N{\times}M)$ planning overhead and $O(N)$ execution I/O regression.}
\label{fig:celebrity}
\end{figure*}

Figure~\ref{fig:celebrity} illustrates the layout.

\textbf{Write amplification on settlement.} When \texttt{settled\_at} transitions
from NULL to a timestamp, PostgreSQL must move the row from the default partition
to the correct time-range partition. This is internally a \texttt{DELETE} from
the default partition followed by an \texttt{INSERT} into the target partition,
updating $M$ indexes on each - comparable in index cost to Approach~A's
cross-table movement, but now imposed automatically by the database engine on
every settlement event.

\textbf{Irreversibility.} Declarative partitioning in PostgreSQL is a one-way
structural change. Reverting a partitioned table to an unpartitioned one requires
creating a new table, bulk-copying all rows, rebuilding all indexes, and swapping
the table - a multi-hour operation at billion-row scale, requiring a maintenance
window. This makes Approach~B a high-risk choice: if its performance impact is
worse than expected, the remediation is itself disruptive.

\begin{table*}[!t]
\centering
\caption{Approach comparison. Approach~C (Purge Job) is selected.}
\label{tab:comparison}
\scriptsize
\begin{tabular}{@{}>{\raggedright\arraybackslash}p{2.2cm}>{\raggedright\arraybackslash}p{2.1cm}>{\raggedright\arraybackslash}p{2.1cm}>{\raggedright\arraybackslash}p{2.1cm}@{}}
\toprule
\textbf{Property}
  & \textbf{Approach A}\newline(Two Tables)
  & \textbf{Approach B}\newline(Partition)
  & \textbf{Approach C}\newline(Purge Job) \\
\midrule
Write amplification
  & $2\times$ on settlement
  & $2\times$ on settlement
  & None \\[4pt]
Index overhead
  & $2\times$ (two tables)
  & $O(N{\times}M)$ planning, $O(N)$ execution
  & Unchanged \\[4pt]
Schema change
  & Yes (new table)
  & Yes (partitioning)
  & No \\[4pt]
Reversibility
  & Moderate
  & Low (one-way)
  & High (stop job) \\[4pt]
Operational complexity
  & High
  & Medium
  & Low \\[4pt]
Partition pruning benefit
  & N/A
  & None (queries without\newline partition key filter)
  & N/A \\
\bottomrule
\end{tabular}
\end{table*}

\subsection{Approach C: Unsettled-Only Table with Periodic Purge (Selected)}

The third approach inverts the archival goal: rather than moving settled data to
a separate store, the primary cluster retains only active transactions. After
a transaction is settled, it remains in the table for a short archival window
(during which it may still be needed for in-flight operations) and is then
deleted by a scheduled purge job. Settled transactions are queryable from the
warm cluster, which continuously replicates all events via CDC and retains
deleted rows.

This approach avoids the write amplification of Approach~A (no cross-table
movement on settlement), the celebrity partition problem of Approach~B (no
partitioning), and requires no schema change. The table structure is identical
to the current transactions table, making the change fully backward-compatible.
The purge job is controllable: it can be paused, rate-limited, and resumed
independently of the application.

\section{Solution Architecture}
\label{sec:architecture}

\subsection{Overview}

The deployed architecture maintains a single transactions table on the primary
cluster, containing only active (and recently settled) transactions. A CDC pipeline continuously replicates all events to the warm cluster; the
archival window $W$ ensures all events are replicated long before they could be
purged. A scheduled
purge job deletes settled rows older than the archival window during off-peak
hours. The application layer routes reads to the appropriate cluster based on
the query's data requirements.

\begin{figure*}[!t]
\centering
\resizebox{0.55\textwidth}{!}{%
\begin{tikzpicture}[
  node distance  = 1.0cm and 1.6cm,
  box/.style     = {draw, rounded corners=3pt, minimum width=4.2cm,
                    minimum height=0.75cm, align=center, font=\footnotesize},
  cluster/.style = {draw, rounded corners=4pt, fill=gray!12,
                    minimum width=7.8cm, minimum height=1.6cm,
                    align=center, font=\footnotesize\bfseries},
  circ/.style    = {draw, circle, minimum size=1.0cm, align=center,
                    font=\footnotesize\bfseries},
  arr/.style     = {-{Stealth[length=5pt]}, thick},
  darr/.style    = {{Stealth[length=5pt]}-{Stealth[length=5pt]}, thick},
  lbl/.style     = {font=\footnotesize, midway}
]

\node[circ] (event) {Payment\\Event};

\node[box, fill=orange!20, below left=1.1cm and 0.5cm of event] (unsettled)
  {UNSETTLED\\{\footnotesize \texttt{settled\_at} IS NULL}};

\node[box, fill=purple!10, left=1.4cm of unsettled] (eval) {Evaluation\\\& Matching};

\node[box, fill=blue!15, below=1.1cm of unsettled] (settled)
  {SETTLED\\{\footnotesize \texttt{settled\_at} = timestamp}};

\node[box, fill=red!15, below=1.1cm of settled] (purge)
  {Purge Job (DEL)};

\node[cluster, fill=orange!10, below=1.8cm of purge] (hot)
  {HOT CLUSTER (Primary)\\[2pt]
   {\normalfont\footnotesize $\sim$6\% of transactions (unsettled) $\leftarrow$ writes}\\
   {\normalfont\footnotesize recently-settled rows $\leftarrow$ within window $W$}};

\node[cluster, fill=blue!10, below=1.4cm of hot] (warm)
  {WARM CLUSTER (Historical)\\[2pt]
   {\normalfont\footnotesize All transactions, including purged settled rows}\\
   {\normalfont\footnotesize $\leftarrow$ reconciliation, audit, fallback reads}};

\draw[arr] (event) -- (unsettled);
\draw[darr] (unsettled) -- node[lbl, above]{domain updates} (eval);
\draw[arr] (unsettled) -- node[lbl, right]{settled} (settled);
\draw[arr] (settled)   -- node[lbl, right]{after archival window $W$} (purge);
\draw[arr] (purge)     -- node[lbl, right]{\footnotesize batch DELETE} (hot);
\draw[arr] (hot)       -- node[lbl, right]{\footnotesize CDC replication (suppresses DELETE)} (warm);

\node[draw, dashed, rounded corners, below=0.6cm of warm,
      minimum width=7.8cm, align=center, font=\footnotesize] (ann) {
  $\sim$94\% of transactions: settled $\to$ warm only\\
  $\sim$6\% of transactions: unsettled $\to$ hot (primary)
};

\end{tikzpicture}%
}
\caption{Transaction lifecycle and hot/warm storage routing. Unsettled
transactions remain in the hot cluster; settled rows are purged after window
$W$ (with CDC suppressing the DELETE so warm retains all history).}
\label{fig:lifecycle}
\end{figure*}
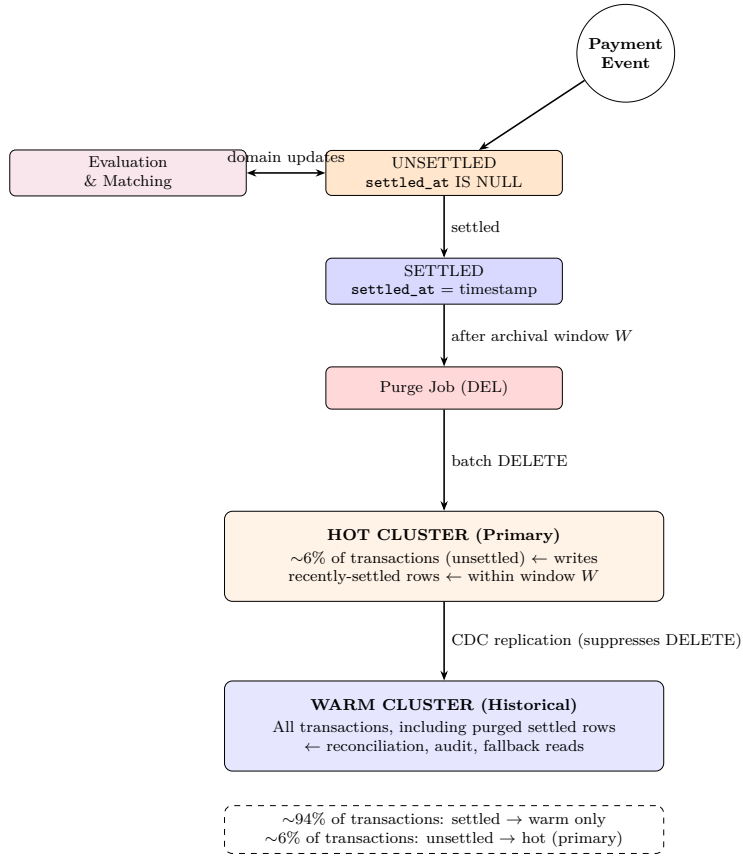

Figure~\ref{fig:lifecycle} illustrates the transaction lifecycle and storage
routing. The hot/warm separation follows established CQRS practice; the
engineering challenge specific to this context is restoring insert safety after
archival breaks primary-key uniqueness guarantees (Section~6).
Migration to the new cluster was performed via logical replication with
reverse replication maintained temporarily for rollback safety. The new cluster
initialized with ${\sim}6\%$ of the original row count.

\subsection{Purge Job}

The purge job deletes rows where \texttt{settled\_at} is older than the archival
window $W$ (on the order of days) plus a clock-skew margin (Section~6.3):
\begin{itemize}[itemsep=0pt, topsep=2pt, parsep=0pt, partopsep=0pt]
\item \textbf{Batched deletes:} configurable batch size limits lock hold time and WAL generation.
\item \textbf{Pause/resume:} pausing retains rows safely; prolonged pauses increase hot-cluster storage above steady state.
\item \textbf{Backpressure:} throttles or pauses when writer CPU/IO exceeds thresholds.
\item \textbf{Off-peak:} runs overnight, sustaining thousands of deletions per second with no impact on concurrent writes.
\end{itemize}

\subsection{Warm Database and CDC Pipeline}

The warm cluster receives all transaction events via
Debezium~\cite{Debezium}, a log-based CDC framework reading PostgreSQL's
logical replication stream. Delete suppression, the key correctness property,
is implemented at the consumer layer: \texttt{DELETE} events are filtered
out, so rows purged from the primary are retained in the warm cluster
indefinitely. No schema changes or triggers are required on the primary.

The CDC pipeline operates with a replication lag of typically under one minute in
steady state, acceptable for settled-data reads that carry no real-time freshness
requirement. Since $W$ (on the order of days) far exceeds this lag, any record eligible for
the deduplication slow path (age ${>}W$) is guaranteed to have been replicated
to the warm cluster well before it could be purged from the hot cluster.
In production, fewer than 0.0005\% of inserts require a warm-cluster lookup
(the deduplication slow path, Section~\ref{sec:dedup}), with observed lookup
p99 latency under 250\,ms. This rate is sensitive
to the choice of $W$: a larger $W$ reduces the fraction of IDs that trigger
warm-cluster lookups; $W$ should be tuned to the
system's event delivery characteristics.
Because logical replication retains unconfirmed changes in its slot
(at-least-once delivery), a CDC outage delays rather than loses data; replication
health is monitored and the purge halts on degradation, and since $W$ far exceeds
replication lag this leaves a multi-day safety margin.

\section{Post-Archival Deduplication}
\label{sec:dedup}

\subsection{The Duplicate Insert Problem}

Archiving records from the primary cluster breaks the uniqueness guarantee of
the primary key. While a record is present in the primary cluster, its primary
key enforces uniqueness: any re-insertion of the same ID will be rejected by the
unique constraint. Once the record is purged, its key is removed from the index.
A subsequent re-insertion of the same ID will succeed, creating a duplicate.

As noted in Section~1.2, duplicate records in a financial system risk
duplicate settlements and monetary loss. Re-insertions occur in practice through
at-least-once event delivery and operational replay mechanisms, both of which
create duplicate risk after archival.

\subsection{Leveraging Monotonic IDs}

The deduplication technique requires the system's IDs to have a recoverable
creation timestamp encoded monotonically in their prefix. At Razorpay, transaction IDs are alphanumeric strings whose prefix encodes a monotonically
increasing timestamp, with a short cryptographic random suffix for collision
resistance. This is structurally equivalent to
ULID~\cite{ULID} and Snowflake IDs~\cite{Twitter10}, both of which embed a
monotonically increasing timestamp in the high-order bits.

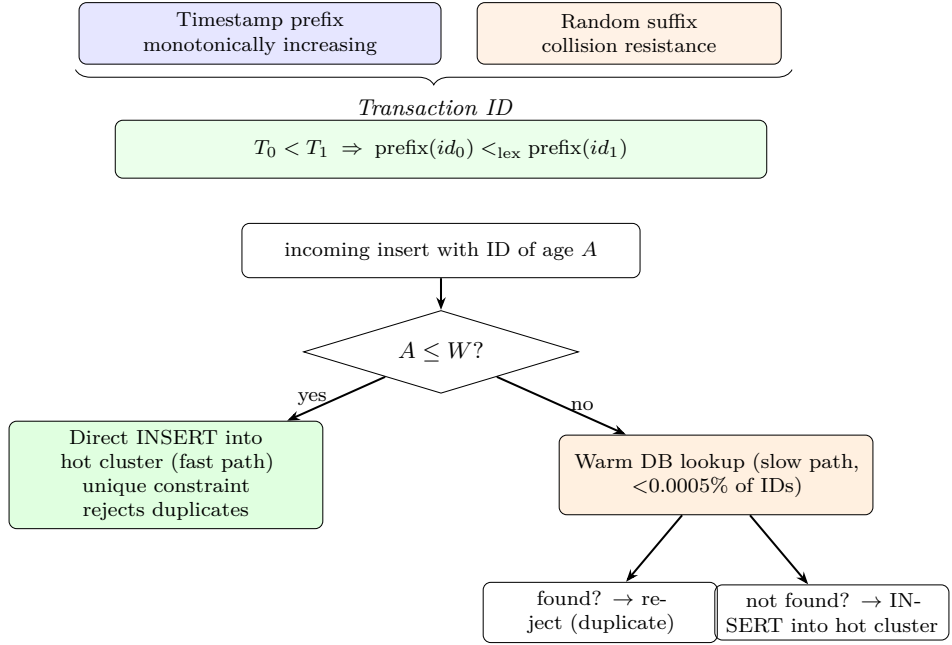
\begin{figure*}[!t]
\centering
\resizebox{0.7\textwidth}{!}{%
\begin{tikzpicture}[
  box/.style={draw, rounded corners=3pt, align=center, font=\footnotesize},
  diam/.style={draw, diamond, aspect=2.2, align=center, font=\small,
               minimum width=3.8cm, minimum height=1.0cm},
  arr/.style={-{Stealth[length=5pt]}, thick},
  lbl/.style={font=\footnotesize, midway}
]

\node[box, fill=blue!10, minimum width=5.0cm, minimum height=0.85cm]
  (ts) at (-2.5, 0) {Timestamp prefix\\monotonically increasing};
\node[box, fill=orange!10, minimum width=4.2cm, minimum height=0.85cm]
  (rnd) at (2.6, 0)  {Random suffix\\collision resistance};

\draw[decorate, decoration={brace, amplitude=6pt, mirror},
      yshift=-2pt]
  (-5.05,-0.43) -- (4.85,-0.43)
  node[midway, below=8pt, font=\small\itshape] {Transaction ID};

\node[box, fill=green!8, minimum width=9.0cm, minimum height=0.80cm]
  (mono) at (0,-1.6)
  {$T_0 < T_1 \;\Rightarrow\; \mathrm{prefix}(id_0) <_{\mathrm{lex}} \mathrm{prefix}(id_1)$};

\node[box, minimum width=5.5cm, minimum height=0.75cm]
  (start) at (0,-3.0)
  {incoming insert with ID of age $A$};

\node[diam] (dec) at (0,-4.4) {$A \leq W$?};

\node[box, fill=green!12, minimum width=4.2cm, minimum height=1.1cm,
      text width=4.1cm]
  (fast) at (-3.8,-6.1)
  {Direct INSERT into\\hot cluster (fast path)\\unique constraint rejects duplicates};

\node[box, fill=orange!12, minimum width=4.2cm, minimum height=1.1cm,
      text width=4.1cm]
  (slow) at (3.8,-6.1)
  {Warm DB lookup (slow path,\\$<$0.0005\% of IDs)};

\node[box, minimum width=3.0cm, minimum height=0.7cm, text width=3.0cm]
  (found)    at (2.2,-8.0) {found? $\to$ reject (duplicate)};
\node[box, minimum width=3.0cm, minimum height=0.7cm, text width=3.0cm]
  (notfound) at (5.4,-8.0) {not found? $\to$ INSERT into hot cluster};

\draw[arr] (start) -- (dec);
\draw[arr] (dec) -- node[lbl, left]{yes} (fast);
\draw[arr] (dec) -- node[lbl, right]{no}  (slow);
\draw[arr] (slow) -- (found);
\draw[arr] (slow) -- (notfound);

\end{tikzpicture}%
}
\caption{Transaction ID structure (top), monotonicity property (middle), and
deduplication routing decision tree (bottom). IDs within the archival window $W$
take the fast path; only older IDs incur a warm DB lookup.}
\label{fig:idstruct}
\end{figure*}

The defining property is: for any two IDs generated at times $T_0 < T_1$, the
prefix of the $T_1$ ID is lexicographically greater than the prefix of the $T_0$
ID. Equivalently, creation time is recoverable from any ID by decoding its
prefix:
\[
  \mathrm{creation\_time}(id)
  = \mathrm{decode\_prefix}(id)
\]
This property - that IDs are temporally ordered and their creation time is
recoverable from the ID itself - is the foundation of the deduplication
technique. Any ID scheme with this property (ULID, Snowflake,
KSUID~\cite{KSUID}) is compatible. Figure~\ref{fig:idstruct} illustrates the ID
structure and deduplication routing.

\subsection{The Deduplication Technique}

Given an archival window $W$, Algorithm~\ref{alg:safe_insert} describes the
insert routing procedure.

\begin{algorithm}[!t]\footnotesize
\DontPrintSemicolon
\SetAlgoLined
\caption{Safe-Insert with ID-Monotonicity Deduplication}
\label{alg:safe_insert}
\KwIn{incoming\_id, record}
\BlankLine
$\mathit{age} \leftarrow \mathrm{current\_time} - \mathrm{creation\_time}(\mathit{incoming\_id})$\;
\BlankLine
\eIf{$\mathit{age} \leq W$}{
  \tcp{ID within archival window; unique constraint handles duplicates}
  \textbf{INSERT} record \textbf{INTO} transactions\;
}{
  \tcp{ID predates window; may have been purged}
  $\mathit{result} \leftarrow$ \textbf{SELECT} id \textbf{FROM} warm\_db
    \textbf{WHERE} id $=$ incoming\_id\;
  \uIf{warm\_db \textbf{unavailable}}{
    \textbf{defer insert} \tcp*{duplicate prevention over throughput}
  }\uElseIf{$\mathit{result}$ \textbf{is not empty}}{
    \textbf{reject insert} \tcp*{duplicate detected}
  }\Else{
    \textbf{INSERT} record \textbf{INTO} transactions\;
  }
}
\end{algorithm}

\textbf{Correctness.} For IDs within the archival window
($\mathit{age} \leq W$): by construction, the purge job only deletes rows where
\texttt{settled\_at} is older than $W$. A transaction settled within the last
$W$ days has \texttt{settled\_at} younger than the purge threshold and is still
present in the primary cluster. A transaction not yet settled is also still
present. Therefore, any ID with age $\leq W$ is guaranteed to be in hot storage
if it exists anywhere, and the unique constraint correctly handles the duplicate
case.

For IDs outside the archival window ($\mathit{age} > W$): the record may
have been settled more than $W$ days ago and purged. The warm cluster, which
retains all records including purged ones, serves as the authoritative check. If
the warm cluster has the record, it is a duplicate; if not, the insert is safe.
If a concurrent insert of the same ID races between the warm-cluster lookup and
the hot-cluster INSERT, the unique constraint on the primary key in the hot
cluster rejects the second insert, preserving safety. The warm-cluster check
could return a false negative only if a record had been purged from the hot
cluster before its insert event replicated to the warm cluster; purge
eligibility requires age greater than $W$ while replication lag is orders of
magnitude smaller than $W$ (Section~5.3), so this window is empty.

\textbf{Warm cluster unavailability.} When the warm cluster is unreachable,
Algorithm~\ref{alg:safe_insert} prioritizes duplicate prevention over insert
throughput: inserts requiring the slow path are deferred until the warm cluster
recovers. This is a deliberate design choice - a re-delivered event can be
retried, whereas a duplicate settlement cannot be undone.

\textbf{Tolerating Clock Skew.} Clock skew between the ID generator and the
insert consumer is assumed negligible relative to $W$. With $W$ on the order of
days, even several hours of drift has no material effect on correctness.
As an additional guard, the purge predicate is strict: rows are deleted
only when \texttt{settled\_at} is older than $W$ plus a skew margin, so the
purge threshold and the fast-path boundary ($\mathit{age} \leq W$) cannot
overlap even under adverse skew.

\textbf{Prefix integrity.} Correctness assumes the decoded prefix never
post-dates true creation time ($\mathrm{decode\_prefix}(id) \leq
\mathrm{creation\_time}$); an insert-time validation gate rejects any ID whose
decoded timestamp lies in the future (beyond skew tolerance), enforcing this
precondition and subsuming the clock-skew assumption above.

\subsection{Edge Case: Decoupled Entity Creation}

The monotonicity assumption holds cleanly when ID generation and record creation
in the downstream system are tightly coupled. However, some entity types exhibit
decoupled creation: the ID is assigned at one point in time (e.g., at initial
request acceptance), while the corresponding downstream record is created only
upon a later confirmation event. If the confirmation occurs more than $W$ days
after the ID was assigned, the incoming ID will have age $> W$ even though the
record has never previously appeared in the downstream system. This triggers a
warm DB lookup (the slow path) even though no prior record exists - a slow-path
misclassification.

The practical impact is bounded, and $W$ should be chosen to minimize this rate.
For example, with $W$ on the order of days, fewer than 0.0005\%
of arriving records have ID age $> W$ - a negligible slow-path rate. Operators
should tune $W$ to drive this rate below an acceptable threshold for their
delivery semantics; the slow-path rate decreases monotonically as $W$ increases.

\subsection{Comparison with Alternatives}

Alternative deduplication approaches include a Redis Bloom filter over purged
IDs~\cite{Bloom70} (tens of GB at this scale, plus an external stateful
dependency) and a separate idempotency table (hundreds of GB, recreating the
storage growth problem). ID-monotonicity requires no additional storage, no
external dependencies, and no ongoing maintenance. The archival window invariant,
already required for correctness, provides the temporal guarantee; the warm DB,
already required for historical reads, doubles as the deduplication authority.

\section{Production Results}
\label{sec:results}

The archival system described in Sections~5 and~6 has been fully deployed to
production and operated continuously since its rollout. Pre-archival baseline metrics were measured
on the source cluster before migration. Post-archival metrics reflect
steady-state operation after full backlog clearance approximately two weeks post-cutover. Storage and row counts are point-in-time measurements from
\texttt{pg\_stat\_user\_tables} and Aurora storage dashboards. Latency metrics
(Section~7.2) are sustained p99 values over a 30-day representative traffic period beginning after the stabilization window.
CPU utilization (Section~7.3) is the sustained writer instance average over the
same period.

Two caveats apply to these comparisons. First, the pre- and post-archival
measurements are separated by the migration period, during which traffic
volume continued to grow; the improvements are not controlled for load
changes, although growing load biases the comparison against the
post-archival system, making the reported reductions conservative. Second,
the analytical figures of Section~4 (e.g., the $N$-fold execution I/O
regression in Figure~\ref{fig:celebrity}) count per-partition operations and
treat each partition-local index scan as comparable to a scan of a single
global index; partition-local indexes are smaller, so those figures are
upper bounds rather than measurements.

\subsection{Storage and Row Count Reduction}

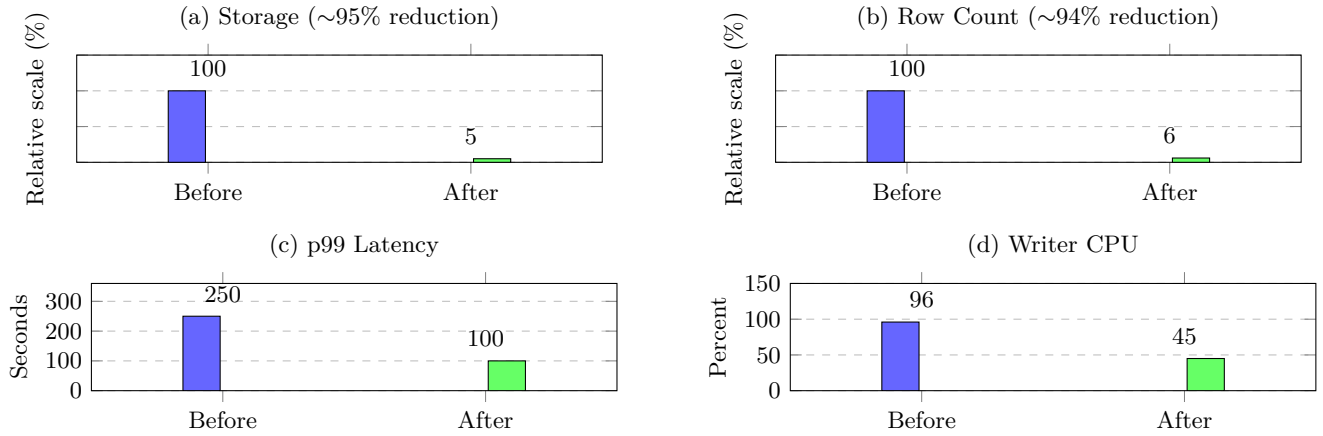
\begin{figure*}[!t]
  \centering
  \begin{subfigure}[b]{0.48\textwidth}
    \centering
    \begin{tikzpicture}
    \begin{axis}[
      width=\linewidth, height=3.0cm,
      ybar, bar width=14pt,
      symbolic x coords={Before,After},
      xtick={Before,After},
      ymin=0, ymax=150,
      ylabel={Relative scale (\%)},
      title={(a) Storage ($\sim$95\% reduction)},
      title style={font=\small},
      ylabel style={font=\small},
      tick label style={font=\small},
      nodes near coords, nodes near coords align={vertical},
      every node near coord/.style={font=\small, yshift=2pt},
      enlarge x limits=0.5,
      ymajorgrids=true, grid style=dashed,
      yticklabels={,,},
    ]
    \addplot[fill=blue!60]  coordinates {(Before,100)};
    \addplot[fill=green!60] coordinates {(After,5)};
    \end{axis}
    \end{tikzpicture}
  \end{subfigure}
  \hfill
  \begin{subfigure}[b]{0.48\textwidth}
    \centering
    \begin{tikzpicture}
    \begin{axis}[
      width=\linewidth, height=3.0cm,
      ybar, bar width=14pt,
      symbolic x coords={Before,After},
      xtick={Before,After},
      ymin=0, ymax=150,
      ylabel={Relative scale (\%)},
      title={(b) Row Count ($\sim$94\% reduction)},
      title style={font=\small},
      ylabel style={font=\small},
      tick label style={font=\small},
      nodes near coords, nodes near coords align={vertical},
      every node near coord/.style={font=\small, yshift=2pt},
      enlarge x limits=0.5,
      ymajorgrids=true, grid style=dashed,
      yticklabels={,,},
    ]
    \addplot[fill=blue!60]  coordinates {(Before,100)};
    \addplot[fill=green!60] coordinates {(After,6)};
    \end{axis}
    \end{tikzpicture}
  \end{subfigure}

  \vspace{0.3em}

  \begin{subfigure}[b]{0.48\textwidth}
    \centering
    \begin{tikzpicture}
    \begin{axis}[
      width=\linewidth, height=3.0cm,
      ybar, bar width=14pt,
      symbolic x coords={Before,After},
      xtick={Before,After},
      ymin=0, ymax=360,
      ylabel={Seconds},
      title={(c) p99 Latency},
      title style={font=\small},
      ylabel style={font=\small},
      tick label style={font=\small},
      nodes near coords, nodes near coords align={vertical},
      every node near coord/.style={font=\small, yshift=2pt},
      enlarge x limits=0.5,
      ymajorgrids=true, grid style=dashed,
    ]
    \addplot[fill=blue!60]  coordinates {(Before,250)};
    \addplot[fill=green!60] coordinates {(After,100)};
    \end{axis}
    \end{tikzpicture}
  \end{subfigure}
  \hfill
  \begin{subfigure}[b]{0.48\textwidth}
    \centering
    \begin{tikzpicture}
    \begin{axis}[
      width=\linewidth, height=3.0cm,
      ybar, bar width=14pt,
      symbolic x coords={Before,After},
      xtick={Before,After},
      ymin=0, ymax=150,
      ylabel={Percent},
      title={(d) Writer CPU},
      title style={font=\small},
      ylabel style={font=\small},
      tick label style={font=\small},
      nodes near coords, nodes near coords align={vertical},
      every node near coord/.style={font=\small, yshift=2pt},
      enlarge x limits=0.5,
      ymajorgrids=true, grid style=dashed,
    ]
    \addplot[fill=blue!60]  coordinates {(Before,96)};
    \addplot[fill=green!60] coordinates {(After,45)};
    \end{axis}
    \end{tikzpicture}
  \end{subfigure}

  \caption{Before/after production metrics on the primary cluster. Blue bars
  show pre-archival values; green bars show post-archival steady-state values.}
  \label{fig:metrics}
\end{figure*}

Table~\ref{tab:beforeafter} summarizes the before-and-after state of the primary
database cluster. The ${\sim}94\%$ reduction in row count and ${\sim}95\%$
reduction in hot storage reflects the removal of the full settled transaction backlog
accumulated over a decade of operation (storage reduces slightly more than row
count due to index overhead elimination). This reduction was realized at
cutover: the replacement cluster was initialized via logical replication with
only unsettled and recently settled rows (Section~5.1), so the new primary
began with roughly 6\% of the original rows; the purge job maintains this
steady state thereafter, rather than shrinking files in place.
Figure~\ref{fig:metrics} presents the key metrics graphically.

\begin{table*}[!t]
\centering\small
\caption{Primary cluster before/after archival deployment.}
\label{tab:beforeafter}
\begin{tabular}{@{}l>{\raggedright\arraybackslash}p{1.8cm}>{\raggedright\arraybackslash}p{2.8cm}l@{}}
\toprule
\textbf{Metric} & \textbf{Pre-Archival} & \textbf{Post-Archival} & \textbf{Reduction} \\
\midrule
Total DB size
  & tens of TB
  & $\sim$5\% of pre-archival size
  & $\sim$95\% \\
Transactions table rows
  & billions
  & $\sim$6\% of pre-archival count
  & $\sim$94\% \\
Settled rows in hot DB
  & $\sim$94\% of total rows
  & only rows settled within the last $W$ days
  & ${>}99\%$ \\
\bottomrule
\end{tabular}
\end{table*}

\subsection{Query Latency}

Service-level p99 processing latency - the primary performance objective - improved
substantially after archival, dropping from approximately
250~seconds pre-archival to approximately 100~seconds post-archival, a ${\sim}60\%$
reduction.

The improvement results from the ${\sim}94\%$ reduction in scan scope combined with
elimination of table bloat. Worst-case merchant settlement query latency dropped
from over 5~minutes to under 1.5~minutes.

\subsection{Writer CPU Utilization}

Writer instance CPU utilization dropped from a sustained 96\% pre-archival to
45\% post-archival - a 51~percentage-point reduction. At 96\%, the writer was
operating near saturation, with minimal headroom to absorb write bursts; at
45\%, the instance has substantial capacity margin for peak load spikes without
approaching throttling or Aurora autoscaling triggers.

Three compounding effects drive the reduction:
\begin{itemize}[itemsep=0pt, topsep=2pt, parsep=0pt, partopsep=0pt]
\item \textbf{Shallower indexes:} ${\sim}94\%$ fewer rows make B-tree indexes smaller and more cache-resident, reducing per-write I/O.
\item \textbf{Smaller scan scope:} settlement queries traverse ${\sim}17{\times}$ less data, proportionally reducing CPU per query.
\item \textbf{Lighter autovacuum:} smaller working set and lower dead-tuple rate yield shorter, less frequent vacuum cycles.
\end{itemize}

\subsection{Vacuum Behavior}

A qualitative but meaningful result of the archival is the change in vacuum
behavior on the primary cluster. Pre-archival, the transactions table received
dead tuples from settlement updates at a rate approaching one per settlement
update on a continuously growing multi-terabyte heap. Post-archival, dead tuples arise
only from updates to the operationally active unsettled subset (${\sim}6\%$ of
the pre-archival row count) - a significantly smaller working set.
Post-migration, \texttt{pg\_stat\_user\_tables} for the transactions table reports
near-zero accumulated dead tuple counts, consistent with the new table starting
with a clean physical layout and receiving a substantially lower ongoing dead
tuple accumulation rate.

\subsection{Infrastructure Cost}

PostgreSQL Aurora cost comprises storage (per GB-month), I/O (per million
requests), and compute (instance hours). The 95\% hot storage size reduction
eliminated most of the storage component, and read I/O dropped with the smaller
working set.

Compute costs were unchanged (writer and reader instances remain sized for peak
write TPS) and the warm cluster added a modest incremental cost. The combined
effect was a 53\% reduction in total monthly infrastructure spend, bounded by the
fixed compute component.

\subsection{Operational Experience}

\textbf{Cutover.} The application was switched during a brief write pause (minutes)
with no user-visible impact. Reverse replication was maintained for $5{\times}W$
as a rollback path. The warm cluster was cut over after the new primary stabilized.

\textbf{Validation.} Consistency was verified via row-level snapshot comparisons
and business-level reconciliation against upstream records. No discrepancies were
found.

\textbf{Dead tuple surge during purge.} Batch \texttt{DELETE} operations
generated dead tuples faster than autovacuum could reclaim. The fix: a targeted
\texttt{VACUUM} scheduled before each purge run, avoiding both global autovacuum
tuning (too aggressive during peak hours) and rate-limiting the purge job (too
slow). Post-fix, dead tuple counts remain near-zero throughout the purge window.

\textbf{Monitoring added post-deployment.} Four metrics were added to the
operational runbook: (1)~\emph{Archival lag alert}: fires if the purge job falls
behind by a configured lag threshold; (2)~\emph{Warm cluster CDC
lag}: alerts on replication lag exceeding a configured threshold;
(3)~\emph{Slow-path insert rate}: baseline ${\sim}0.0005\%$ of inserts (insert deduplication slow-path rate) - a spike
indicates late-arriving event replay or ID generation anomalies; (4)~\emph{Dead
tuple count}: \texttt{pg\_stat\_user\_tables} on the transactions table, alerting
on unexpected accumulation between vacuum cycles.

\section{Related Work}
\label{sec:related}

PostgreSQL partitioning tooling such as pg\_partman~\cite{pgpartman},
TimescaleDB~\cite{Freedman19}, and Citus~\cite{Cubukcu21} assumes the partition
key is present for all rows of interest; the Celebrity Partition Problem arises
when this assumption fails. Hot/cold separation has also been studied as
temperature-based tiering: anti-caching~\cite{DeBrabant13} evicts cold tuples
based on access recency, and Project Siberia~\cite{Levandoski13} classifies
record temperature from access logs. These systems infer coldness from
observed access patterns, and cold data may become hot again. Lifecycle-aware
archival differs in both signal and guarantee: settlement is an explicit,
domain-semantic terminal state that is immediate and irreversible, so settled
rows are guaranteed immutable and can be removed rather than demoted.
Monotonic ID schemes (Snowflake~\cite{Twitter10},
ULID~\cite{ULID}, KSUID~\cite{KSUID}) are well-studied for generation and
sortability, but we are not aware of prior work applying them to
post-archival deduplication. Bloom filters~\cite{Bloom70} and cuckoo
filters~\cite{Fan14} address set membership at scale but introduce external
dependencies; our approach replaces them with a deterministic temporal comparison.
Temporal tables (SQL:2011) and bitemporal models~\cite{ChandraSegev93} version
data in place, compounding rather than resolving storage growth.

\section{Known Limitations}
\label{sec:limitations}

\textbf{Purge job as a single point of operational control.} The correctness of
the deduplication technique depends on the purge job maintaining the archival
window invariant. If the purge job malfunctions and deletes rows prematurely,
the deduplication technique may miss duplicates. Monitoring the purge job lag
and enforcing the archival window constraint in the job's delete predicate
mitigates this risk, but it is an operational dependency that does not exist in
a non-archiving system.

\textbf{Warm cluster as a new dependency.} Reads that previously hit only the
primary cluster now potentially fall through to the warm cluster. The warm
cluster's availability and replication lag become part of the application's
correctness envelope.

\section{Conclusion}
\label{sec:conclusion}

We presented a lifecycle-aware archival system for a billion-row payment
settlements database on PostgreSQL Aurora. The system exploits the sharp
asymmetry between active and terminal rows to retain only the active subset in
hot storage, achieving a 95\% hot storage size reduction, 53\% infrastructure
cost reduction, ${\sim}60\%$ p99 latency improvement, and a 51~percentage-point CPU reduction at
peak write TPS. We additionally characterized the Celebrity Partition Problem and
introduced an ID-monotonicity deduplication technique requiring no external
dependencies. Both contributions apply broadly to lifecycle-asymmetric relational
data.

\section*{Acknowledgements}

We thank Arpit Bhayani, Vivek Agarwal, Arjun Tomer, Parin Katariya, Praveen
Parihar, and Ashwath Kumar for detailed feedback and editorial review; Sudhanshu Sharma
for leading the database cluster rollout and migrations; and the Razorpay
platform teams for production monitoring and operational support.

\balance
{\small
\bibliographystyle{unsrtnat}
\bibliography{references}

@manual{PostgreSQL14,
  title        = {{PostgreSQL} 14 Documentation},
  author       = {{The PostgreSQL Global Development Group}},
  year         = {2021},
  note         = {\url{https://www.postgresql.org/docs/14/}. Accessed: 2026-07-12}
}

@misc{KSUID,
  author       = {{Segment Engineering}},
  title        = {{KSUID}: K-Sortable Globally Unique {IDs}},
  year         = {2017},
  howpublished = {\url{https://github.com/segmentio/ksuid}},
  note         = {Accessed: 2026-07-12}
}

@misc{Debezium,
  author       = {{Red Hat}},
  title        = {Debezium: Change Data Capture for Databases},
  year         = {2016},
  howpublished = {\url{https://debezium.io/}},
  note         = {Accessed: 2026-07-12}
}

@article{Bloom70,
  author       = {Bloom, Burton H.},
  title        = {Space/Time Trade-offs in Hash Coding with Allowable Errors},
  journal      = {Communications of the ACM},
  volume       = {13},
  number       = {7},
  pages        = {422--426},
  year         = {1970},
  doi          = {10.1145/362686.362692},
  publisher    = {ACM}
}

@inproceedings{Fan14,
  author       = {Fan, Bin and Andersen, Dave G. and Kaminsky, Michael and Mitzenmacher, Michael D.},
  title        = {Cuckoo Filter: Practically Better Than {Bloom}},
  booktitle    = {Proceedings of the 10th ACM International Conference on Emerging Networking Experiments and Technologies (CoNEXT)},
  pages        = {75--88},
  year         = {2014},
  publisher    = {ACM},
  doi          = {10.1145/2674005.2674994}
}

@inproceedings{Cubukcu21,
  author       = {{\c{C}}ubuk{\c{c}}u, Umur and Erdogan, Ozgun and Pathak, Sumedh and Sannakkayala, Sudhakar and Slot, Marco},
  title        = {Citus: Distributed {PostgreSQL} for Data-Intensive Applications},
  booktitle    = {Proceedings of the 2021 ACM SIGMOD International Conference on Management of Data},
  pages        = {2490--2502},
  year         = {2021},
  publisher    = {ACM},
  doi          = {10.1145/3448016.3457551}
}

@misc{Freedman19,
  author       = {Freedman, Michael J. and Kulkarni, Ajay},
  title        = {{TimescaleDB}: {SQL} Made Scalable for Time-Series Data},
  year         = {2018},
  howpublished = {\url{https://assets.timescale.com/docs/downloads/tigerdata-whitepaper.pdf}},
  note         = {Technical Report. Accessed: 2026-07-12}
}

@misc{Twitter10,
  author       = {{Twitter Engineering}},
  title        = {Announcing {Snowflake}},
  year         = {2010},
  howpublished = {\url{https://web.archive.org/web/20240422012321/https://blog.twitter.com/engineering/en_us/a/2010/announcing-snowflake}},
  note         = {Accessed: 2026-07-12}
}

@misc{pgpartman,
  author       = {Fiske, Keith},
  title        = {pg\_partman: Partition Management Extension for {PostgreSQL}},
  year         = {2014},
  howpublished = {\url{https://github.com/pgpartman/pg_partman}},
  note         = {Accessed: 2026-07-12}
}

@misc{ULID,
  author       = {Feerasta, Alizain},
  title        = {{ULID}: Universally Unique Lexicographically Sortable Identifier},
  year         = {2016},
  howpublished = {\url{https://github.com/ulid/spec}},
  note         = {Accessed: 2026-07-12}
}

@article{DeBrabant13,
  author       = {DeBrabant, Justin and Pavlo, Andrew and Tu, Stephen and Stonebraker, Michael and Zdonik, Stan},
  title        = {Anti-Caching: A New Approach to Database Management System Architecture},
  journal      = {Proceedings of the VLDB Endowment},
  volume       = {6},
  number       = {14},
  pages        = {1942--1953},
  year         = {2013},
  doi          = {10.14778/2556549.2556575}
}

@inproceedings{Levandoski13,
  author       = {Levandoski, Justin J. and Larson, Per-{\AA}ke and Stoica, Radu},
  title        = {Identifying Hot and Cold Data in Main-Memory Databases},
  booktitle    = {Proceedings of the 29th IEEE International Conference on Data Engineering (ICDE)},
  pages        = {26--37},
  year         = {2013},
  publisher    = {IEEE},
  doi          = {10.1109/ICDE.2013.6544811}
}

@inproceedings{ChandraSegev93,
  author       = {Chandra, Rakesh and Segev, Arie},
  title        = {Managing Temporal Financial Data in an Extensible Database},
  booktitle    = {Proceedings of the 19th International Conference on Very Large Data Bases (VLDB)},
  pages        = {302--313},
  year         = {1993},
  publisher    = {Morgan Kaufmann}
}
}

\end{document}